\documentclass[notitlepage,prd,preprintnumbers,longbibliography,showpacs]{revtex4-1}
\usepackage[%
  colorlinks=true,
  urlcolor=blue,
  linkcolor=blue,
  citecolor=blue
]{hyperref}
\usepackage{slashed,color,amsmath,amssymb,mathrsfs}
\usepackage{graphicx}
\usepackage{hyperref}
\usepackage{longtable}
\usepackage{array}
\usepackage{accents}
\usepackage{tensor}
\usepackage{autobreak}

\allowdisplaybreaks

\newcommand{\la}{\langle}
\newcommand{\ra}{\rangle}
\newcommand{\chip}{\chi_+}
\newcommand{\chim}{\chi_-}
\newcommand{\fp}{f_+}
\newcommand{\fm}{f_-}

\newcommand{\gamf}{\gamma_5}
\newcommand{\fgamma}{\gamf\gamma}

\newcommand*{\chid}{\chi^\dag}
\def\FL{F_L}
\def\FR{F_R}

\newcommand{\Pd}{P^{\dag}}

\newcommand{\Ps}{{P^*}}
\newcommand{\Psd}{{P^{*\dag}}}

\newcommand{\Hb}{\bar{H}}

\newcommand{\deltas}{\delta^*}

\begin{document}
\title{Chiral Lagrangians for mesons with a single heavy quark}
\author{Shao-Zhou Jiang$^{1}$}\email{jsz@gxu.edu.cn}
\author{Yan-Rui Liu$^2$}\email{yrliu@sdu.edu.cn}
\author{Qin-He Yang$^1$}

\affiliation{$^1$Key Laboratory for Relativistic Astrophysics, Department of Physics, Guangxi University, Nanning 530004, People's Republic of China\\
$^2$School of Physics, Shandong University, Jinan 250100, People's Republic of China
}
\date{\today}

\begin{abstract}
We construct the relativistic chiral Lagrangians for heavy-light mesons $(Q\bar{q})$ to the $\mathcal{O}(p^4)$ order. From $\mathcal{O}(p^2)$ to $\mathcal{O}(p^4)$, there are 17, 67, and 404 independent terms in the flavor $SU(2)$ case and 20, 84, and 655 independent terms in the flavor $SU(3)$ case. The Lagrangians in the heavy quark limit are also obtained. From $\mathcal{O}(p^2)$ to $\mathcal{O}(p^4)$, there are 7, 25, and 136 independent terms in the flavor $SU(2)$ case and 8, 33, and 212 independent terms in the flavor $SU(3)$ case. The relations between low-energy constants based on the heavy quark symmetry are also given up to the $\mathcal{O}(p^3)$ order.
\end{abstract}
%\pacs{}
\maketitle
%\tableofcontents

%%%%%%%%%%%%%%%%%%%%%%%%%%%%%%%%%%%%%%%%%%%%%%%%%%%%%%
\section{Introduction}\label{intr}
%%%%%%%%%%%%%%%%%%%%%%%%%%%%%%%%%%%%%%%%%%%%%%%%%%%%%%

The spontaneous breaking of the global chiral symmetry of QCD is an important feature in the low-energy nonperturbative region of strong interactions. It has been widely accepted that the low-lying pseudoscalar mesons are those Goldstone bosons generated from the symmetry breaking. The effective theory based on this symmetry and its breaking is the chiral perturbation theory (ChPT) \cite{Weinberg:1978kz,Gasser:1983yg,Gasser:1984gg}, which originally describes only low-energy dynamics of such mesons. Later, the theory was extended to cases involving octet baryons \cite{Krause:1990xc}, decuplet baryons \cite{Jenkins:1991es,Hemmert:1997ye}, and heavy quark hadrons \cite{Yan:1992gz,Wise:1992hn}. The matter fields involved in the present study are those heavy-light mesons whose quark content is $Q\bar{q}$ ($Q=c,b; q=u,d,s$).

Because of the light quark, the low-energy interactions for the heavy-light mesons are governed by the chiral symmetry. In addition, the interactions also obey the spin-flavor heavy quark (HQ) symmetry in the limit $M_Q\to\infty$ \cite{Wise:1992hn,Isgur:1989vq,Isgur:1989ed,Georgi:1990um}. The heavy quark flavor symmetry means that different heavy flavors have the same dynamics while the heavy quark spin symmetry results in degenerate hadron doublets containing states with different spins. In ChPT, increasing number of low-energy constants (LECs) need to be determined when high order chiral corrections are considered. For the case involving the heavy-light mesons, the heavy quark symmetry may provide relations between LECs \cite{Burdman:1992gh}. Before we can determine their values with other approaches, these constraints just from symmetry are certainly instructive. Of course, the corrections to such relations due to finite quark mass may also be needed for more detailed investigations.

With the chiral Lagrangians involving heavy-light mesons, a wide range of problems can be studied \cite{Guo:2008gp,Guo:2009ct,Altenbuchinger:2013vwa,Liu:2012vd,Yao:2015qia,Guo:2015dha,Wang:2014jua,Du:2016tgp,Sun:2016tmz,Xu:2017tsr,Cheng:2017oqh,Du:2017ttu,Guo:2018kno}, such as properties of heavy-light mesons, mass difference between heavy-light mesons in the same doublet, interactions between the Goldstone bosons and the heavy-light mesons, interactions between heavy-light mesons, properties of new open-flavor particles \cite{Chen:2016spr}, and so on. Up to now, the chiral Lagrangian in the sector of light pseudoscalar mesons has been constructed up to the $\mathcal{O}(p^8)$ order \cite{Fearing:1994ga,HerreraSiklody:1996pm,Bijnens:1999sh,Bijnens:2001bb,Ebertshauser:2001nj,Cata:2007ns,Haefeli:2007ty,Jiang:2014via,Bijnens:2018lez}. Recently, there are also developments in the sector of light baryons \cite{Fettes:2000gb,Oller:2006yh,Frink:2006hx,Jiang:2016vax,Jiang:2017yda,Jiang:2018mzd,Holmberg:2018dtv}. However, the existing heavy-light meson chiral Lagrangian is still at low orders. The leading order result was obtained long time ago \cite{Burdman:1992gh,Yan:1992gz,Wise:1992hn}. For higher order results, only parts of them were constructed for special problems, which can be found in Refs. \cite{Guo:2008gp,Guo:2009ct,Altenbuchinger:2013vwa,Liu:2012vd,Yao:2015qia,Guo:2015dha,Wang:2014jua,Du:2016tgp,Sun:2016tmz,Xu:2017tsr,Cheng:2017oqh,Du:2017ttu,Guo:2018kno}. Some similar works are about the $SU(2)$ pion-kaon chiral Lagrangian. This chiral Lagrangian has the same structures as that for the heavy-light pseudoscalar mesons. It has been constructed to the $\mathcal{O}(p^4)$ order \cite{Roessl:1999iu,Du:2016xbh,Du:2016ntw}. In the present work, we systematically construct the relativistic chiral Lagrangians in the sector of heavy-light mesons up to the fourth chiral order. To find some relations of LECs by using the heavy quark symmetry, we also construct directly chiral Lagrangians with the superfield $H$ containing the $J^P=0^-$ and $1^-$ $Q\bar{q}$ mesons. By comparing the relativistic Lagrangians with those in the HQ limit, one obtains relations between LECs.

This paper is organized as follows. In Sec. \ref{rev0}, we review briefly the building blocks for the construction of chiral Lagrangians. In Sec. \ref{construction}, from the structures of Lagrangians to linear relations between various ingredients, we introduce the procedure to construct the heavy-light meson chiral Lagrangians step by step. In Sec. \ref{relation}, the way to find relations between LECs with the heavy quark symmetry is introduced. In Sec. \ref{results}, we list our results. The last Sec. \ref{summ} is a short summary.

\section{Definitions and Building Blocks}\label{rev0}

In this section, we give the building blocks necessary for the construction of Lagrangians. Some simple properties are also shown. These building blocks involve both relativistic and HQ forms. One may find details about them in Refs. \cite{Gasser:1984gg,Gasser:1987rb,Fearing:1994ga,Bijnens:1999sh,Bijnens:2001bb,Cata:2007ns,Jiang:2014via,Bijnens:2018lez,Yan:1992gz,Burdman:1992gh,Wise:1992hn,Wise:1993wa,Jenkins:1992zx,Casalbuoni:1996pg}.

\subsection{Goldstone bosons and external sources}\label{rev}

The QCD Lagrangian $\mathscr{L}^0_{\mathrm{QCD}}$ with $N_f$-flavour massless quarks is
\begin{eqnarray}
\mathscr{L}=\mathscr{L}^0_{\mathrm{QCD}}+\bar{q}(\slashed{v}+\slashed{a}\gamma_5-s+ip\gamma_5)q,
\end{eqnarray}
where $q$ denotes the light quark field. $s$, $p$, $v^\mu$, and $a^\mu$ are scalar, pseudoscalar, vector, and axial-vector external sources, respectively. The tensor source and the $\theta$ term are ignored in this paper. As usual, both $a^\mu$ and $v^\mu$ are considered traceless in the flavor $SU(3)$ case, but only $a^\mu$ is traceless in the flavor $SU(2)$ case to study the electroweak interactions.

In ChPT, the low-lying pseudoscalar mesons are considered to be Goldstone bosons coming from the spontaneous breaking of the global symmetry $SU(N_f)_L\times SU(N_f)_R$ into $SU(N_f)_V$. The meson field $u$ in matrix form transforms as
\begin{align}
u\to g_R uh^\dag=h ug_L^\dag
\end{align}
under the chiral rotation, where $g_L$ and $g_R$ are group elements of $SU(N_f)_L$ and $SU(N_f)_R$, respectively, and $h$ is a compensator field which is a function of the pion fields.

Usually, the meson fields and external sources are combined to building blocks whose forms are as follows,
\begin{align}\label{df}
\begin{split}
u^\mu&=i\{u^\dag(\partial^\mu-ir^\mu)u-u(\partial^\mu-il^\mu)u^\dag\},\\
\chi_\pm&=u^\dag\chi u^\dag\pm u\chi^\dag u,\\
h^{\mu\nu}&=\nabla^\mu u^\nu+\nabla^\nu u^\mu,\\
f_+^{\mu\nu}&=u F_L^{\mu\nu} u^\dag+ u^\dag F_R^{\mu\nu} u,\\
f_-^{\mu\nu}&=u F_L^{\mu\nu} u^\dag- u^\dag F_R^{\mu\nu} u=-\nabla^\mu u^\nu+\nabla^\nu u^\mu,
\end{split}
\end{align}
where $r^\mu=v^\mu+a^\mu$, $l^\mu=v^\mu-a^\mu$, $\chi=2B_0(s+ip)$, $F_R^{\mu\nu}=\partial^{\mu}r^{\nu}-\partial^{\nu}r^{\mu}-i[r^{\mu},r^{\nu}]$, $F_L^{\mu\nu}=\partial^{\mu}l^{\nu}-\partial^{\nu}l^{\mu}-i[l^{\mu},l^{\nu}]$, and $B_0$ is a constant related to the quark condensate. The covariant derivative for any building block $X$ is defined through
\begin{align}
\nabla^\mu X&=\partial^\mu X+[\Gamma^\mu,X],\\
\Gamma^\mu&=\frac{1}{2}\{u^\dag(\partial^\mu-ir^\mu)u+u(\partial^\mu-il^\mu)u^\dag\}.
\end{align}
The advantage of using these building blocks is that all of them transform under the chiral rotation ($R$) in the same way,
\begin{align}
X\xrightarrow{R} X'=hXh^\dag.
\end{align}

\subsection{Heavy-light mesons}\label{rev1}

A heavy-light meson contains one heavy quark $Q$ ($c$ or $b$) and one light antiquark $\bar{q}$ ($\bar{u}$, $\bar d$, or $\bar s$). The lowest lying heavy-light mesons are the pseudoscalar $P$ with $J^P=0^-$ and vector $P^*$ with $J^P=1^-$. In the flavour $SU(3)$ case ($Q$ as a flavor singlet), they are represented as row vectors,
\begin{align}
P =\left\{\begin{array}{ll}
(D^0,\;D^+,\;D^+_s),\\
(\bar{B}^-,\;\bar{B}^0,\;\bar{B}^0_s),
\end{array}\right.
\quad
P^* =\left\{\begin{array}{ll}
(D^{*0},\;D^{*+},\;D^{*+}_s),\\
(\bar{B}^{*-},\;\bar{B}^{*0},\;\bar{B}^{*0}_s).
\end{array}\right.
\end{align}
In the flavor $SU(2)$ case, the third $Q\bar{s}$ mesons need to be removed. The Lagrangians in these two cases have different independent chiral-invariant terms and we consider results in both cases. The covariant derivative for $\tilde{P}$ ($P$ or $P^{*\mu}$) is
\begin{align}
D^\mu \tilde{P}^\dag&=(\partial^\mu+\Gamma^\mu)\tilde{P}^\dag,\\
D^\mu \tilde{P}&=\tilde{P}(\overleftarrow{\partial}^\mu+\Gamma^{\mu\dag}),\\
D^{\mu\nu\cdots\rho}&\equiv\frac{1}{n!}(\underbrace{D^{\mu}D^{\nu}\cdots D^{\rho}}_{n}+\text{full permutation of $D$'s}).\label{nD}
\end{align}
Eq. \eqref{nD} defines a totally symmetrical covariant derivative like the $\pi N$ case \cite{Fettes:2000gb}. The reason for this definition is that permutations of derivatives acting on a building block do not change the chiral dimension (as Eq. \eqref{nn} below). The defined symmetrical derivative will simplify some calculations (see Sec. \ref{relation}). The chiral transformations for heavy-light meson fields read
\begin{align}
\tilde{P}\xrightarrow{R}\tilde{P}'=\tilde{P}h^\dag,\quad\tilde{P}^\dag\xrightarrow{R}\tilde{P}^{\prime\dag}=h\tilde{P}^\dag.
\end{align}

To adopt the heavy quark symmetry, one collects the heavy-light mesons in a superfield as usual \cite{Georgi:1991mr},
\begin{align}
H=\sqrt{M}\frac{1+\slashed{v}}{2}(P^*_{Q,\mu}\gamma^\mu+\delta P_{Q}\gamf),\quad
\Hb\equiv\gamma^0 H^\dag\gamma^0,\label{defH}
\end{align}
where $M\equiv M_P=M_{P^*}$ is the heavy-light meson masses in the HQ limit and $v^\mu$ with $v^2=1$ is the velocity of heavy-light mesons, $P^*_{Q,\mu}$ and $P_{Q}$ only contain the annihilation operator. Now, $H$ contains only annihilation operators for $Q\bar{q}$ mesons and its mass dimension is 3/2. We here use $\delta$ to denote the arbitrary relative phase between the mesons $P_Q$ and $P^*_Q$. Different conventions exist in the literature, e.g. $\delta=1$ in \cite{Yan:1992gz}, $\delta=-1$ in \cite{Wise:1993wa}, and $\delta=i$ in \cite{Manohar:2000dt}. This phase does not have physical effects and its choice does not impact on the form of Lagrangians, either. Scaling the superfield by $e^{-iM v\cdot x}$ will modify the energy measure from $M_Q$ to $M$ and the covariant derivative on matter fields becomes $D^\mu H(x)=-iMv^\mu H(x)$. Obviously, the chiral transformations for $H$ and $\bar{H}$ are the same as those for $\tilde{P}$ and $\tilde{P}^\dag$, respectively.

\section{Lagrangian Construction}\label{construction}

This section shows the basic steps to construct the Lagrangian for heavy-light mesons. First, one analyzes the structures of chiral Lagrangians because they have effects on some properties of building blocks. Secondly, one establishes the $P$-parity, $C$-parity, and Hermitian properties of all the building blocks. Thirdly, one finds out available linear relations in order to reduce linearly dependent terms. Finally, one constructs all possible structures of the chiral Lagrangian and gets independent terms by using the linear relations.

\subsection{Structures of chiral Lagrangians}
The relativistic heavy-light meson chiral Lagrangian can be written as
\begin{align}
\mathscr{L}&=\mathscr{L}_{PP}+\mathscr{L}_{P^*P^*}+\mathscr{L}_{PP^*}\\
&=\sum_n C_n P\cdots\Pd+\sum_m C_m \Ps\cdots\Psd+\sum_p C_p (P\cdots\Psd+\Ps\cdots\Pd),\label{rLform}
\end{align}
where $\mathscr{L}_{PP}$, $\mathscr{L}_{P^*P^*}$, and $\mathscr{L}_{PP^*}$ represent the interaction terms involving only heavy pseudoscalar mesons, only heavy vector mesons, and both heavy pseudoscalar and heavy vector mesons, respectively. The symbol ``$\cdots$'' includes allowed combinations of building blocks given in Sec. \ref{rev} and appropriate coefficients ($\pm 1$ or $\pm i$)  to keep the symmetry of $\mathscr{L}$. For convenience, the LECs ($C_n$, $C_m$, and $C_p$) are all assumed to be real constants and we use the convention that all the possible covariant derivatives in ``$\cdots$'' act on the right side heavy-light meson fields.% We also choose the convention $\delta=1$ in presenting our final results.

To find out relations between LECs in the HQ limit, we also construct chiral Lagrangians involving the superfield $H$ directly. The Lagrangian in this formalism looks like
\begin{align}
\mathscr{L}=\sum_n D_n\la H\cdots\Gamma\Hb\ra,\label{hLform}
\end{align}
where $D_n$'s represent LECs in this case, $\Gamma$ is an element of the Clifford algebra, and $\la\cdots\ra$ means trace in the spin space. If flavour traces for building blocks are needed in ``$\cdots$'', we also use this symbol $\la\cdots\ra$. The heavy quark symmetry requires that the position of $\Gamma$ should be after $H$ but before $\Hb$.

\subsection{Properties of building blocks}\label{bbp}

The properties of the building blocks have been discussed in a lot of references. Here we only collect relevant results. One may find details about them in Refs. \cite{Gasser:1983yg,Gasser:1984gg,Bijnens:1999sh,Fearing:1994ga,Bijnens:2001bb,Ebertshauser:2001nj,Jiang:2014via,Bijnens:2018lez,Fettes:2000gb,Manohar:2000dt,Wise:1993wa,Yan:1992gz}.

Table \ref{blbt} lists the chiral dimensions, parity transformations ($P$), charge conjugation transformations ($C$), and Hermitian transformations (h.c.) of the building blocks, the matter fields, and the Levi-Civita tensor. Since the heavy-light mesons are not purely neutral states, the phases for the charge conjugation transformation of them are uncertain. Choosing ``+'' for $P$ is natural since $J^{PC}=0^{--}$ are exotic quantum numbers. For $P^*$, we use the convention ``$+$'' and will discuss another one.

\begin{table*}[!h]
\caption{\label{blbt}Chiral dimension (Dim), parity ($P$), charge conjugation ($C$), and Hermiticity (h.c.) of the building blocks, the matter fields, and the Levi-Civita tensor.}
%\begin{ruledtabular}
{%\renewcommand\arraystretch{1.3}
\begin{tabular}{ccccc}
	\hline\hline
	                   & Dim &         $P$         &            $C$             &          h.c.           \\
	\hline
	    $u^{\mu}$      &  1  &     $-u_{\mu}$      &       $(u^{\mu})^T$        &        $u^{\mu}$        \\
	   $h^{\mu\nu}$    &  2  &    $-h_{\mu\nu}$    &      $(h^{\mu\nu})^T$      &      $h^{\mu\nu}$       \\
	   $\chi_{\pm}$    &  2  &   $\pm\chi_{\pm}$   &      $(\chi_{\pm})^T$      &    $\pm \chi_{\pm}$     \\
	$f_{\pm}^{\mu\nu}$ &  2  & $\pm f_{\pm\mu\nu}$ & $\mp (f_{\pm}^{\mu\nu})^T$ &   $ f_{\pm}^{\mu\nu}$   \\
	       $P$         &  0  &        $-P$         &        $(P^\dag)^T$        &        $P^\dag$         \\
	    $P^{*\mu}$     &  0  &    $P^{*}_{\mu}$    &     $(P^{*\mu\dag})^T$     &     $P^{*\mu\dag}$      \\
	    $D^\mu P$      &  0  &     $-D_\mu P$      &     $(D^\mu P^\dag)^T$     &    $(D^\mu P)^\dag$     \\
	 $D^\mu P^{*\nu}$  &  0  &  $D_\mu P^{*}_\nu$  &  $(D^\mu P^{*\nu\dag})^T$  & $(D^\mu P^{*\nu})^\dag$ \\
   $\varepsilon^{\mu\nu\lambda\rho}$ & 0 & $-\varepsilon_{\mu\nu\lambda\rho}$ & $\varepsilon^{\mu\nu\lambda\rho}$ & $\varepsilon^{\mu\nu\lambda\rho}$\\
	\hline\hline
\end{tabular}
}
%\end{ruledtabular}
\end{table*}

Table \ref{cabt} lists the corresponding properties of the Clifford algebra and the velocity of heavy-light mesons, which are considered between $H$ and $\Hb$ as \eqref{hLform}. $H$, $\Hb$, and $v^\mu$ are chiral dimensionless and their properties are considered together with Clifford algebra, like the $\pi N$ case in Ref. \cite{Fettes:2000gb}. Table \ref{cabt} only displays the extra signs. We do not show anything about $\gamf$ because $\langle H\gamf\Hb\rangle=0$ gives no contributions. $H$ only contains the $Q\bar{q}$ fields, but not $\bar{Q}q$ fields. Hence the Lagrangian in heavy quark symmetry does not have to be $C$-invariant. The meaning of the charge conjugation in Table \ref{cabt} will be discussed in Section \ref{relation}.

\begin{table*}[!h]
\caption{\label{cabt}Chiral dimension (Dim), parity ($P$), charge conjugation ($C$), and Hermiticity (h.c.) of the Clifford algebra elements and the velocity of heavy-light mesons.}
%\begin{ruledtabular}
\begin{tabular}{ccccc}
	\hline\hline
	                  & Dim & $P$ & $C$ & h.c. \\
	\hline
	       $1$        &  0  & $+$ & $+$ & $+$  \\
	 $\gamma^{\mu}$   &  0  & $+$ & $-$ & $+$  \\
	 $\fgamma^{\mu}$  &  0  & $-$ & $+$ & $+$  \\
	$\sigma^{\mu\nu}$ &  0  & $+$ & $-$ & $+$  \\
	    $v^{\mu}$     &  0  & $+$ & $-$ & $+$  \\
	\hline\hline
\end{tabular}
%\end{ruledtabular}
\end{table*}

\subsection{Linear relations}\label{lr}

Linear relations exist which are essential in reducing the chiral-invariant terms to a minimal set. For details about them, one may consult Refs. \cite{Bijnens:1999sh,Fearing:1994ga,Bijnens:2001bb,Jiang:2014via,Bijnens:2018lez}.

\begin{enumerate}
\item \label{pirl}Partial integration.

Ignoring higher order terms, one has
\begin{align}
0\doteq\tilde{P}\accentset{\leftharpoonup}{D}^\mu X\tilde{P}^\dag+\tilde{P} XD^\mu\tilde{P}^\dag,\label{par}
\end{align}
where $X$ is any building block or their products and ``$\doteq$'' means that both sides are approximately equal with their difference appearing at the order $\mathcal{O}(p^1)$. With this relation, we can move the covariant derivatives to the right position so that they act only on the field $\tilde{P}^\dag$.

\item Schouten identity.

This is a relation about the Levi-Civita tensor,
\begin{align}
\epsilon^{\mu\nu\lambda\rho}A^\sigma-\epsilon^{\sigma\nu\lambda\rho}A^\mu-\epsilon^{\mu\sigma\lambda\rho}A^\nu
-\epsilon^{\mu\nu\sigma\rho}A^\lambda-\epsilon^{\mu\nu\lambda\sigma}A^\rho=0,\label{sir}
\end{align}
where $A$ is anything having Lorentz index (indices). The five indices in the left-hand side are totally antisymmetric.

\item Equations of motions (EOMs).

The EOMs and subsidiary condition for light pseudoscalar and heavy-light mesons are
\begin{align}
\nabla_\mu u^\mu&\doteq\frac{i}{2}\bigg(\chim-\frac{1}{N_f}\la\chim\ra\bigg),\label{eomb}\\
(D^2+M_P^2)P^\dag&\doteq 0,\label{eoms}\\
(D^2+M_{P^*}^2)P_{ \mu}^{*\dag}&\doteq 0,\\
D^\mu P_{\mu}^{*\dag}&\doteq 0,\label{eomp}\\
v^\mu P_{Q,\mu}^{*\dag}&=0,\label{eomh}
\end{align}
where $N_f$ is the number of light quark flavours and the conjugations of these equations are omitted. Eq. \eqref{eomh} only works in the heavy quark limit. The right-hand sides of Eqs. \eqref{eoms}--\eqref{eomp} are at least at the order $\mathcal{O}(p^1)$. At the $\mathcal{O}(p^1)$ order, they contain one $u^\mu$. Hence, $D^2\tilde{P}$ can be changed to $-M^2_{\tilde{p}}\tilde{P}$ and does not happen in the Lagrangian. $D^\mu P^*_{\mu}$ is at the order $\mathcal{O}(p^1)$. It removes the redundant degree of freedom of $P_{\mu}^{*}$ field, and $D^\mu P^*_{\mu}$ does not appear in the Lagrangian, either.

\item Covariant derivatives and Bianchi identity.

The commutative relation for the covariant derivatives acting on any building block $X$ is
\begin{align}
&[\nabla^\mu,\nabla^\nu]X=[\Gamma^{\mu\nu},X],\label{nn}\\
&\Gamma^{\mu\nu}=\frac{1}{4}[u^\mu,u^\nu]-\frac{i}{2}\fp^{\mu\nu}.
\end{align}
Rewriting it explicitly, one has
\begin{align}
\nabla^\mu \nabla^\nu X-\nabla^\nu \nabla^\mu X
-\frac{1}{4}u^\mu u^\nu X+\frac{1}{4}u^\nu u^\mu X+\frac{i}{2}\fp^{\mu\nu}X
+\frac{1}{4}X u^\mu u^\nu-\frac{1}{4}X u^\nu u^\mu -\frac{i}{2}X\fp^{\mu\nu}=0.
\end{align}
Another relation about covariant derivatives is Bianchi identity
\begin{align}
&\nabla^{\mu}\Gamma^{\nu\lambda}+\nabla^{\nu}\Gamma^{\lambda\mu}+\nabla^{\lambda}\Gamma^{\mu\nu}=0.
\end{align}
Its explicit form is
\begin{align}
\nabla^\mu\fp^{\nu\lambda}+\nabla^\nu\fp^{\lambda\mu}+\nabla^\lambda\fp^{\mu\nu}
+\frac{i}{2}[u^{\mu},\fm^{\nu\lambda}]+\frac{i}{2}[u^{\nu},\fm^{\lambda\mu}]+\frac{i}{2}[u^{\lambda},\fm^{\mu\nu}]=0.
\end{align}
These two explicit relations are for determining the strict relations of LECs which will be discussed in Sec. \ref{relation}.

\item Cayley-Hamilton relations.

Any $2\times 2$ matrices $A$ and $B$ have the relation
\begin{align}
AB+BA-A\la B\ra-B\la A\ra-\la AB\ra+\la A\ra\la B\ra=0.\label{ch2}
\end{align}
Any $3\times 3$ matrices $A$, $B$, and $C$ have the relation
\begin{align}
0={}&ABC+ACB+BAC+BCA+CAB+CBA-AB\la C\ra-AC\la B\ra-BA\la C\ra-BC\la A\ra-CA\la B\ra\notag\\
&-CB\la A\ra-A\la BC\ra-B\la AC\ra-C\la AB\ra-\la ABC\ra-\la ACB\ra+A\la B\ra\la C\ra+B\la A\ra\la C\ra+C\la A\ra\la B\ra\notag\\
&+\la A\ra\la BC\ra+\la B\ra\la AC\ra+\la C\ra\la AB\ra-\la A\ra\la B\ra\la C\ra.\label{ch3}
\end{align}

\item Contact terms.

The contact terms need to be picked up independently. Such terms appear only at the $\mathcal{O}(p^4)$ order. To show their irrelevance with pion fields, we change the building blocks from $u^\mu$, $h^{\mu\nu}$, $f_\pm^{\mu\nu}$, and $\chi_\pm$ to $F_{R,L}^{\mu\nu}$, $\chi$, and $\chi^\dag$. The relevant relations are
\begin{align}
F_L^{\mu\nu}&=\frac{1}{2}u^\dag(f_+^{\mu\nu}+f_-^{\mu\nu})u,\\
F_R^{\mu\nu}&=\frac{1}{2}u(f_+^{\mu\nu}-f_-^{\mu\nu})u^\dag,\\
\chi&=\frac{1}{2}u(\chi_++\chi_-)u,\\
\chi^\dag&=\frac{1}{2}u^\dag(\chi_+-\chi_-)u^\dag.
\end{align}
We show the properties of these building blocks (LR-basis) \cite{Fearing:1994ga,Jiang:2014via} in Table \ref{lrbhb}.
The number of resultant contact terms is found small. They are listed at the end of each part for $\mathscr{L}_{PP}$, $\mathscr{L}_{P^*P^*}$, and $\mathscr{L}_{PP^*}$ in Table \ref{rp4} and such terms in the HQ limit are given at the end of Table \ref{hp4}.
\begin{table*}[h]
\caption{\label{lrbhb}Chiral rotations (R), parity ($P$), charge conjugation ($C$), and Hermiticity (h.c.) of the LR-basis.}
\begin{tabular}{ccccc}
\hline\hline  & $R$ & $P$ & $C$ & h.c. \\
\hline
$\chi$ & $g_R\chi g_L^\dag$ & $\chid$ & $\chi^T$ & $\chid$\\
$\chid$ & $g_L\chid g_R^\dag$ & $\chi$ & $\chi^{\dag T}$ & $\chi$\\
$\FL^{\mu\nu}$ & $g_L\FL^{\mu\nu}g_L^\dag$ & $\FR^{\mu\nu}$ & $-(\FR^{\mu\nu})^T$ & $\FL^{\mu\nu}$\\
$\FR^{\mu\nu}$ & $g_R\FR^{\mu\nu}g_R^\dag$ & $\FL^{\mu\nu}$ &  $-(\FL^{\mu\nu})^T$ & $\FR^{\mu\nu}$\\
\hline\hline
\end{tabular}
\end{table*}
\end{enumerate}

The process to pick up independent terms is very boring and is done by computer. The details about the operation have been presented in Refs. \cite{Jiang:2014via,Jiang:2016vax,Jiang:2017yda}.

\section{LEC relations in the heavy quark limit}\label{relation}

According to the heavy quark symmetry, relations exist among LECs for $PP^\dag$ terms, those for $P^*P^{*\dag}$ terms, and those for $PP^{*\dag}$ terms. In order to find them, we firstly redefine the independent terms and their corresponding LECs in Eq. \eqref{rLform} to be
\begin{align}
\tilde{O}_n=O_n/M^r,\quad\tilde{C}_n=C_n M^r,
\end{align}
where $r$ is the number of covariant derivative acting on the heavy-light meson fields. Now, all $\tilde{C}_n$'s at a given order have the same mass dimension.% In this paper, we ignore the $O(1/M)$ contribution.

At least two methods can be used to get the LEC relations. The first one is to change the relativistic Lagrangians to the HQ form. With Eq. \eqref{defH}, one obtains
\begin{align}
\sqrt{M}P_Q=\frac{1}{2}\deltas\la H\gamf\ra,\;\sqrt{M}P^*_{Q,\mu}=\frac{1}{2}\la H\gamma_\mu\ra,\;
\sqrt{M}P_Q^\dag=-\frac{1}{2}\delta\la\Hb\gamf\ra,\;\sqrt{M}P^{*\dag}_{Q,\mu}=\frac{1}{2}\la\Hb\gamma_\mu\ra. \label{PHr}
\end{align}
These fields contain only operators to annihilate or generate $Q\bar{q}$ mesons, but no operators for $\bar{Q}q$ mesons. If we assume that the fields in Eq. \eqref{rLform} also describe only $Q\bar{q}$ mesons, the Lagrangian there can be changed to that in Eq. \eqref{hLform} by using the Fierz identity. Retaining only terms satisfying the heavy quark symmetry and comparing independent terms, one can obtain relations between $C_n$'s and $D_n$'s.% Some terms which break heavy quark symmetry give the constraint conditions of LECs.

The second method is opposite by changing the form of Eq. \eqref{hLform} to that of Eq. \eqref{rLform},
\begin{align}
\la H\Hb\ra\to{}&M(-2 P_Q P_Q^{\dag}+2 P_Q^{*\mu}P^{*\dag}_{Q,\mu}),\\
\la H\gamf\Hb\ra\to{}&0,\label{Hgamf}\\
\la H\gamma^\mu\Hb\ra\to{}&M(-\la Hv^\mu\Hb\ra)=\sqrt{M}(-2i P_Q D^\mu P_Q^{\dag}/M+2i P_Q^{*\nu}D^\mu P^{*\dag}_{Q,\nu}/M),\\
\la H\gamf\gamma^{\mu}\Hb\ra\to{}&M(-2\varepsilon^{\mu\nu\lambda\rho}P^{*}_{Q,\lambda}D_{\nu}P^{*\dag}_{Q,\rho}/M
+2\delta P_Q P_Q^{*\dag\mu}
+2\delta^{*} P_Q^{*\mu} P_Q^{\dag}),\\
\la H\sigma^{\mu\nu}\Hb\ra\to{}&M(2i P_Q^{*\mu}P_Q^{*\dag\nu}
-2i P_Q^{*\nu}P_Q^{*\dag\mu}
+2i\varepsilon^{\mu\nu\lambda\rho}\delta P_Q D_{\lambda}P^{*\dag}_{Q,\rho}/M
+2i\varepsilon^{\mu\nu\lambda\rho}\delta^{*}P^{*}_{Q,\rho} D_{\lambda}P_Q^{\dag}/M),
\end{align}
where we have used the definition $\la\gamma^\mu\gamma^\nu\gamma^\lambda\gamma^\rho\gamf\ra=-4i\varepsilon^{\mu\nu\lambda\rho}$. The factor $M$ comes from the definition in Eq. \eqref{defH}. From the above equations, one finds that only structures $\la H\Hb\ra$,  $\la H\gamf\gamma^{\mu}\Hb\ra$, and $\la H\sigma^{\mu\nu}\Hb\ra$ exist in the final results, a feature consistent with the pion-nucleon case \cite{Fettes:2000gb}. In order to obtain the relativistic Lagrangian, the right-hand sides of the above equations also need $C$ invariant. If one substitutes $\tilde{P}_Q\to\tilde{P}$ and chooses the ``$C$-parity'' of the Clifford algebra and the velocity as those in Table \ref{cabt}, these terms automatically contain the $C$-invariant.

To get the exact relations between $\tilde{C}_k$ and $D_l$, the strict linear relations are needed. In Sec. \ref{lr}, we have found them in the relativistic case. Hence, we choose the second method to do the calculation. This method also avoids complex calculation from the Fierz identity. The relations are
\begin{align}
\tilde{C}_k=M\sum_{l}D_lA_{lk},\label{rcd}
\end{align}
where $M$ is a usual normalization factor coming from Eq. \eqref{defH}. All elements in matrix $A_{lk}$ are dimensionless. Since the number of $D_l$ is much less than the number of $\tilde{C}_k$ (see the results in Sec. \ref{results}), $D_l$ may be obtained more easily in other ways. If all $D_l$ are known, Eq. \eqref{rcd} gives a rough estimation of $\tilde{C}_k$. It also gives some constraint conditions of $\tilde{C}_k$ in the heavy quark limit.

To calculate Eq. \eqref{rcd}, we avoid the approximate relations (marked by ``$\doteq$'') in Sec. \ref{lr} as far as possible. Higher order contribution of the EOM for pseudoscalar mesons (Eq. \eqref{eomb}) does not work to the $\mathcal{O}(p^2)$ order, and higher order contribution of the EOMs for heavy-light mesons does not work to the $\mathcal{O}(p^3)$ order. Hence, all relations in Sec. \ref{lr} are strict ones to the $\mathcal{O}(p^3)$ order.

\section{Results}\label{results}

Following the above steps, we get the final results expressed as
\begin{align}
&\mathscr{L}^{(m)}=\sum_{n} C^{(m)}_n O^{(m)}_n=\sum_{n} \tilde{C}^{(m)}_n \tilde{O}^{(m)}_n,\;N_f=3\\
&\mathscr{L}^{(m)}=\sum_{n} c^{(m)}_n o^{(m)}_n=\sum_{n} \tilde{c}^{(m)}_n \tilde{o}^{(m)}_n,\;N_f=2\\
&\mathscr{L}^{(m)}_{\mathrm HQ}=\sum_{n} D^{(m)}_n P^{(m)}_n,\;N_f=3\\
&\mathscr{L}^{(m)}_{\mathrm HQ}=\sum_{n} d^{(m)}_n p^{(m)}_n,\;N_f=2.
\end{align}
where $m$ is the chiral dimension.

\subsection{Results at the $\mathcal{O}(p^1)$ and $\mathcal{O}(p^2)$ orders}

The obtained relativistic result at the leading order,
\begin{align}
\mathscr{L}^{(1)}={}&D_\mu P D^\mu\Pd-M_P^2 P\Pd-\frac{1}{2}(D^{\mu}P^{*\nu}-D^{\nu}P^{*\mu})(D_{\mu}P^{*\dag}_{\nu}-D_{\nu}P^{*\dag}_{\mu})+M^2_{P^*}P^{*\mu}P^{*\dag}_\mu\notag\\
&+\frac{1}{2}f_Q(P u^{\mu}\Psd_\mu+\text{H.c.})
+\frac{1}{4}g_Q\varepsilon^{\mu\nu\lambda\rho}(P^{*}_{\rho} u_{\lambda}(D_{\mu}P^{*\dag}_{\nu}-D_{\nu}P^{*\dag}_{\mu})+\text{H.c.}),
\end{align}
is the same as that in Ref. \cite{Yan:1992gz}. The form obeying the heavy quark symmetry is \cite{Wise:1992hn}
\begin{align}
\mathscr{L}^{(1)}_{\mathrm HQ}={}&\la Hiv^\mu D_{\mu}\Hb\ra-\frac{1}{2}g\la H u_\lambda\gamf\gamma^\lambda\Hb\ra.
\end{align}
The relations between $f_Q$, $g_Q$, and $g$ are found to be
\begin{align}
f_Q=2g_Q M=-2gM.
\end{align}
The results at this order are applicable for both two- and three- flavour cases.

\begin{table*}[htbp]
\caption{\label{rp2}The $\mathcal{O}(p^2)$ order relativistic results. The columns 2, 3, and 4 (5, 6, and 7) are for the flavor $SU(2)$ (SU(3)) case. When a term $O_n$ is not given a label in the 2nd (5th) column, it is not independent and can be expressed with terms having a label in the 2nd (5th) column. ``I'' means that the structures of those terms are chosen as independent ones in the HQ limit.}
\begin{ruledtabular}
\begin{tabular}{lcrrcrr}
$O_n$ & $SU(2)$ & $\tilde{c}^{(2)}_n$ & $\tilde{c}^{(2)}_n$ & $SU(3)$ & $\tilde{C}^{(2)}_n$ & $\tilde{C}^{(2)}_n$\\
\hline%\addlinespace[2pt]
$P u^{\mu}u_{\mu}P^{\dag}$ & 1 & $-2d^{(2)}_{1}$ & $\mathrm{I}\quad$ & 1 & $-2D^{(2)}_{1}$  & $\mathrm{I}\quad$  \\
$P u^{\mu}u^{\nu}D_{\mu\nu}P^{\dag}$ & 2 & $2d^{(2)}_{2}$ & $\mathrm{I}\quad$ & 2 & $2D^{(2)}_{2}$  & $\mathrm{I}\quad$  \\
$P\la u^{\mu}u_{\mu}\ra P^{\dag}$ &  &  &  & 3 & $-2D^{(2)}_{4}$  & $\mathrm{I}\quad$  \\
$P\la u^{\mu}u^{\nu}\ra D_{\mu\nu}P^{\dag}$ &  &  &  & 4 & $2D^{(2)}_{5}$  & $\mathrm{I}\quad$  \\
$P\chip P^{\dag}$ & 3 & $-2d^{(2)}_{6}$ & $\mathrm{I}\quad$ & 5 & $-2D^{(2)}_{7}$  & $\mathrm{I}\quad$  \\
$P\la\chip\ra P^{\dag}$ & 4 & $-2d^{(2)}_{7}$ & $\mathrm{I}\quad$ & 6 & $-2D^{(2)}_{8}$  & $\mathrm{I}\quad$  \\
$P^{*\mu}u_{\mu}u^{\nu}{P^{*\dag}}_{\nu}$ & 5 & $-2d^{(2)}_{3}$ & $\mathrm{I}\quad$ & 7 & $-2D^{(2)}_{3}$  & $\mathrm{I}\quad$  \\
$P^{*\mu}u^{\nu}u_{\mu}{P^{*\dag}}_{\nu}$ & 6 & $2d^{(2)}_{3}$ & $-\tilde{c}^{(2)}_{5}$ & 8 & $2D^{(2)}_{3}$  & $-\tilde{C}^{(2)}_{7}$  \\
$P^{*\mu}u^{\nu}u_{\nu}{P^{*\dag}}_{\mu}$ & 7 & $2d^{(2)}_{1}$ & $-\tilde{c}^{(2)}_{1}$ & 9 & $2D^{(2)}_{1}$  & $-\tilde{C}^{(2)}_{1}$  \\
$P^{*\mu}u^{\nu}u^{\lambda}D_{\nu\lambda}{P^{*\dag}}_{\mu}$ & 8 & $-2d^{(2)}_{2}$ & $-\tilde{c}^{(2)}_{2}$ & 10 & $-2D^{(2)}_{2}$  & $-\tilde{C}^{(2)}_{2}$  \\
$P^{*\mu}\la u_{\mu}u^{\nu}\ra{P^{*\dag}}_{\nu}$ &  &  &  & 11 & $0\quad$  & $0\quad$  \\
$P^{*\mu}\la u^{\nu}u_{\nu}\ra{P^{*\dag}}_{\mu}$ &  &  &  & 12 & $2D^{(2)}_{4}$  & $-\tilde{C}^{(2)}_{3}$  \\
$P^{*\mu}\la u^{\nu}u^{\lambda}\ra D_{\nu\lambda}{P^{*\dag}}_{\mu}$ &  &  &  & 13 & $-2D^{(2)}_{5}$  & $-\tilde{C}^{(2)}_{4}$  \\
$iP^{*\mu}{f_{+\mu}}^{\nu}{P^{*\dag}}_{\nu}$ & 9 & $4d^{(2)}_{4}$ & $\mathrm{I}\quad$ & 14 & $4D^{(2)}_{6}$  & $\mathrm{I}\quad$  \\
$iP^{*\mu}\la{f_{+\mu}}^{\nu}\ra{P^{*\dag}}_{\nu}$ & 10 & $4d^{(2)}_{5}$ & $\mathrm{I}\quad$ &  &  &  \\
$P^{*\mu}\chip{P^{*\dag}}_{\mu}$ & 11 & $2d^{(2)}_{6}$ & $-\tilde{c}^{(2)}_{3}$ & 15 & $2D^{(2)}_{7}$  & $-\tilde{C}^{(2)}_{5}$  \\
$P^{*\mu}\la\chip\ra{P^{*\dag}}_{\mu}$ & 12 & $2d^{(2)}_{7}$ & $-\tilde{c}^{(2)}_{4}$ & 16 & $2D^{(2)}_{8}$  & $-\tilde{C}^{(2)}_{6}$  \\
$\varepsilon^{\mu\nu\lambda\rho}P u_{\mu}u_{\nu}D_{\lambda}{P^{*\dag}}_{\rho}+\mathrm{H.c.}$ & 13 & $-2d^{(2)}_{3}$ & $\tilde{c}^{(2)}_{5}$ & 17 & $-2D^{(2)}_{3}$  & $\tilde{C}^{(2)}_{7}$  \\
$P{f_{-}}^{\mu\nu}D_{\mu}{P^{*\dag}}_{\nu}+\mathrm{H.c.}$ & 14 & $0\quad$ & $0\quad$ & 18 & $0\quad$  & $0\quad$  \\
$P h^{\mu\nu}D_{\mu}{P^{*\dag}}_{\nu}+\mathrm{H.c.}$ & 15 & $0\quad$ & $0\quad$ & 19 & $0\quad$  & $0\quad$  \\
$i\varepsilon^{\mu\nu\lambda\rho}P f_{+\mu\nu}D_{\lambda}{P^{*\dag}}_{\rho}+\mathrm{H.c.}$ & 16 & $2d^{(2)}_{4}$ & $\frac{1}{2}\tilde{c}^{(2)}_{9}$ & 20 & $2D^{(2)}_{6}$  & $\frac{1}{2}\tilde{C}^{(2)}_{14}$  \\
$i\varepsilon^{\mu\nu\lambda\rho}P\la f_{+\mu\nu}\ra D_{\lambda}{P^{*\dag}}_{\rho}+\mathrm{H.c.}$ & 17 & $2d^{(2)}_{5}$ & $\frac{1}{2}\tilde{c}^{(2)}_{10}$ &  &  &  \\
%\hline
\end{tabular}
\end{ruledtabular}
\end{table*}

We show the $\mathcal{O}(p^2)$ Lagrangian in the relativistic form and HQ form in Tables \ref{rp2} and \ref{hp2}, respectively. The 2nd and 5th columns of Table \ref{rp2} (2nd and 3rd columns of Table \ref{hp2}) give the labels for each term in the flavor $SU(2)$ case and $SU(3)$ case, respectively. The 3rd and 6th columns of Table \ref{rp2} list the corresponding LECs in the HQ limit. The 4th and 7th columns of Table \ref{rp2} display the LEC relations between the relativistic terms according to the heavy quark symmetry, where ``I'' means that corresponding terms can be treated as the independent ones. Some monomials only happen in the either $SU(2)$ or $SU(3)$ case because of the Cayley-Hamilton relations and the convention of the trace of the vector source $v^\mu$. Hence, the other column is not labelled. Only a few analogous results in the references are found. The $\mathcal{O}(p^2)$ order $\mathscr{L}_{PP}$ is the same as that in Ref. \cite{Du:2016tgp}.

\begin{table*}[!h]
\caption{\label{hp2}The $\mathcal{O}(p^2)$ order results in the HQ limit. When a term $P_n$ is not given a label in the 2nd (3rd) column, it is not independent and can be expressed with terms having a label in the 2nd (3rd) column.}
%\begin{ruledtabular}
\begin{tabular}{lcc}
\hline\hline $P_n$ & $SU(2)$ & $SU(3)$\\
\hline%\addlinespace[2pt]
$\la H u^{\mu}u_{\mu}\Hb\ra$ & 1 & 1  \\
$\la H u^{\mu}u^{\nu}v_{\mu}v_{\nu}\Hb\ra$ & 2 & 2  \\
$i\la H u^{\mu}u^{\nu}\sigma_{\mu\nu}\Hb\ra$ & 3 & 3  \\
$\la H\la u^{\mu}u_{\mu}\ra\Hb\ra$ &  & 4  \\
$\la H\la u^{\mu}u^{\nu}\ra v_{\mu}v_{\nu}\Hb\ra$ &  & 5  \\
$\la H{f_{+}}^{\mu\nu}\sigma_{\mu\nu}\Hb\ra$ & 4 & 6  \\
$\la H\la{f_{+}}^{\mu\nu}\ra\sigma_{\mu\nu}\Hb\ra$ & 5 &  \\
$\la H\chip\Hb\ra$ & 6 & 7  \\
$\la H\la\chip\ra\Hb\ra$ & 7 & 8  \\
\hline\hline
\end{tabular}
%\end{ruledtabular}
\end{table*}

\subsection{Results at the $\mathcal{O}(p^3)$ and $\mathcal{O}(p^4)$ orders}

The $\mathcal{O}(p^3)$ and $\mathcal{O}(p^4)$ order results are too long and we give them in Appendix \ref{app1}. The relativistic results are collected in Tables \ref{rp3} and \ref{rp4} while those in the HQ limit are listed in Tables \ref{hp3} and \ref{hp4}. The 3rd and 6th columns of Table \ref{rp3} show the corresponding LECs in the HQ limit while the 4th and 7th columns of the same table display the LEC relations between the relativistic terms obtained from the heavy quark symmetry. ``I'' in Tables \ref{rp3} and \ref{rp4} again means that the relevant terms are considered independent in the HQ limit. Some long expressions marked with ``*'' in Table \ref{rp3} are given explicitly below the table.

At present, we are not able to give the strict LEC relations for terms at the $\mathcal{O}(p^4)$ order. The $\mathcal{O}(p^1)$ order EOMs will appear because of the Schouten identity. Schouten identity can change the positions of some indexes and will give the factors as $D^\mu\Psd_\mu$ or $D^2\tilde{P}$. Hence the LECs at the $\mathcal{O}(p^1)$ order will appear in these relations. The exact relations need the determination of the inverse of a large symbolic matrix. Hence, we only mark the independent terms in the HQ limit in Table \ref{rp4}. In the table, the 52--57, 241--253, and 402--404 terms in the two-flavour case (97-99, 413--418, and 655 terms in the three-flavour case) are contact terms. In Table \ref{hp4}, the 128--136 terms in the two-flavour case (209--212 terms in the three-flavour case) are contact terms.

\subsection{Discussions}
We have chosen the convention $\delta=1$ in presenting our final results. If one wants to use another convention, all the results need not be changed. For the $C$-parity transformation of $P^*$, we also use the ``$+$'' convention. Another convention only has an impact on $\mathscr{L}_{PP^*}$. Let us consider any $C$-, $P$- and h.c.-invariant $\mathscr{L}_{PP^*}$ term
\begin{align}
(P O_{\mu}\Psd^{\mu}+\delta_C\Ps^{\mu} O_{C,\mu} \Pd),\label{cphc}
\end{align}
where $O_{\mu}$ is any possible structure, $O_{C,\mu}$ is an appropriate structure keeping the symmetry, and $\delta_C$ is the $C$-parity transformation factor of $P^*$. If one chooses an opposite sign of $\delta_C$, an extra $i$ factor is needed to keep Hermiticity.

%%%%%%%%%%%%%%%%%%%%%%%%%%%%%%%%%%%%%%%%%%%%%
\section{Summary}\label{summ}
%%%%%%%%%%%%%%%%%%%%%%%%%%%%%%%%%%%%%%%%%%%%%

In the present paper, we extend our previous studies and construct the relativistic chiral Lagrangians for mesons with a heavy quark to one loop, both for the flavor $SU(3)$ case and for the flavor $SU(2)$ case. The chiral Lagrangians in the heavy quark limit are also obtained. The number of independent terms in the heavy quark limit is much less than that in the relativistic case, which is illustrated in Table \ref{numberofterms}. By comparing independent terms in the relativistic form and those in the HQ limit, one finds LEC relations at each order which result from the heavy quark symmetry. These relations would get corrections once the breaking of heavy quark symmetry is considered.

\begin{table}[htbp]
\caption{Number of independent terms at each chiral order.}\label{numberofterms}
\begin{tabular}{ccccc|cccc}\hline\hline
&&\multicolumn{2}{c}{Relativistic}&&&\multicolumn{2}{c}{HQ limit}\\
Chiral order&${\cal O}(p^1)$&${\cal O}(p^2)$&${\cal O}(p^3)$&${\cal O}(p^4)$&${\cal O}(p^1)$&${\cal O}(p^2)$&${\cal O}(p^3)$&${\cal O}(p^4)$\\
\hline$SU(2)$&1&17&67&404&1&7&25&136\\
$SU(3)$&1&20&84&655&1&8&33&212\\\hline\hline
\end{tabular}
\end{table}

%%%%%%%%%%%%%%%%%%%%%%%%%%%%%%%%%%%%%%%
\section*{Acknowledgements}
%%%%%%%%%%%%%%%%%%%%%%%%%%%%%%%%%%%%%%%

This work was supported by the National Science Foundation of China (NSFC) under Grant Nos. 11565004 and 11775132, Guangxi Science Foundation under Grant Nos. 2018GXNSFAA281180 and 2017AD22006, the special funding for Guangxi distinguished professors (Bagui Yingcai and Bagui Xuezhe) and High Level Innovation Team and Outstanding Scholar Program in Guangxi Colleges.

\appendix
\section{$\mathcal{O}(p^3)$ and $\mathcal{O}(p^4)$ order results}\label{app1}

\begin{longtable}[!h]{lcrrcrr}
\caption{The $\mathcal{O}(p^3)$ order relativistic results. The columns 2, 3, and 4 (5, 6, and 7) are for the flavor $SU(2)$ (SU(3)) case. When a term $O_n$ is not given a label in the 2nd (5th) column, it is not independent and can be expressed with terms having a label in the 2nd (5th) column. ``I'' means that the structures of those terms are chosen as independent ones in the HQ limit. Relations marked with ``*'' are given below this table.}\label{rp3}\\
\hline\hline $O_n$ & $SU(2)$ & $\tilde{c}^{(3)}_n$ & $\tilde{c}^{(3)}_n$ & $SU(3)$ & $\tilde{C}^{(3)}_n$ & $\tilde{C}^{(3)}_n$\\
\hline\endfirsthead

\hline\hline $O_n$ & $SU(2)$ & $\tilde{c}^{(3)}_n$ & $\tilde{c}^{(3)}_n$ & $SU(3)$ & $\tilde{C}^{(3)}_n$ & $\tilde{C}^{(3)}_n$\\
\hline\endhead

\hline % \multicolumn{8}{c}{Continued on next page}\\
\endfoot

\hline\endlastfoot

$\varepsilon^{\mu\nu\lambda\rho}P u_{\mu}u_{\nu}u_{\lambda}D_{\rho}P^{\dag}$ & 1 & $-2d^{(3)}_{5}$ & $\mathrm{I}\quad$ & 1 & $-2D^{(3)}_{9}$  & $\mathrm{I}\quad$  \\
$\varepsilon^{\mu\nu\lambda\rho}P\la u_{\mu}u_{\nu}u_{\lambda}\ra D_{\rho}P^{\dag}$ &  &  &  & 2 & $-2D^{(3)}_{10}$  & $\mathrm{I}\quad$  \\
$P u^{\mu}{f_{-\mu}}^{\nu}D_{\nu}P^{\dag}+\mathrm{H.c.}$ & 2 & $-2d^{(3)}_{17}$ & $\mathrm{I}\quad$ & 3 & $-2D^{(3)}_{24}$  & $\mathrm{I}\quad$  \\
$P u^{\mu}{h_{\mu}}^{\nu}D_{\nu}P^{\dag}+\mathrm{H.c.}$ & 3 & $-2d^{(3)}_{20}$ & $\mathrm{I}\quad$ & 4 & $-2D^{(3)}_{27}$  & $\mathrm{I}\quad$  \\
$P u^{\mu}h^{\nu\lambda}D_{\mu\nu\lambda}P^{\dag}+\mathrm{H.c.}$ & 4 & $2d^{(3)}_{21}$ & $\mathrm{I}\quad$ & 5 & $2D^{(3)}_{28}$  & $\mathrm{I}\quad$  \\
$i\varepsilon^{\mu\nu\lambda\rho}P f_{+\mu\nu}u_{\lambda}D_{\rho}P^{\dag}+\mathrm{H.c.}$ & 5 & $2d^{(3)}_{12}$ & $\mathrm{I}\quad$ & 6 & $2D^{(3)}_{18}$  & $\mathrm{I}\quad$  \\
$i\varepsilon^{\mu\nu\lambda\rho}P\la f_{+\mu\nu}\ra u_{\lambda}D_{\rho}P^{\dag}$ & 6 & $2d^{(3)}_{13}$ & $\mathrm{I}\quad$ &  &  &  \\
$i\varepsilon^{\mu\nu\lambda\rho}P\la f_{+\mu\nu}u_{\lambda}\ra D_{\rho}P^{\dag}$ &  &  &  & 7 & $2D^{(3)}_{19}$  & $\mathrm{I}\quad$  \\
$iP\nabla^{\mu}{f_{+\mu}}^{\nu}D_{\nu}P^{\dag}$ & 7 & $2d^{(3)}_{23}$ & $\mathrm{I}\quad$ & 8 & $2D^{(3)}_{33}$  & $\mathrm{I}\quad$  \\
$iP\la\nabla^{\mu}{f_{+\mu}}^{\nu}\ra D_{\nu}P^{\dag}$ & 8 & $2d^{(3)}_{24}$ & $\mathrm{I}\quad$ &  &  &  \\
$iP u^{\mu}\chim D_{\mu}P^{\dag}+\mathrm{H.c.}$ & 9 & $2d^{(3)}_{16}$ & $\mathrm{I}\quad$ & 9 & $2D^{(3)}_{23}$  & $\mathrm{I}\quad$  \\
$\varepsilon^{\mu\nu\lambda\rho}{P^{*}}_{\mu}u_{\nu}u_{\lambda}u^{\sigma}D_{\rho}{P^{*\dag}}_{\sigma}+\mathrm{H.c.}$ & 10 & $-2d^{(3)}_{2}$ & $\mathrm{I}\quad$ & 10 & $*\quad$  & $\mathrm{I}\quad$  \\
$\varepsilon^{\mu\nu\lambda\rho}{P^{*}}_{\mu}u_{\nu}u_{\lambda}u^{\sigma}D_{\sigma}{P^{*\dag}}_{\rho}+\mathrm{H.c.}$ & 11 & $d^{(3)}_{2}+d^{(3)}_{4}$ & $\mathrm{I}\quad$ & 11 & $*\quad$  & $\mathrm{I}\quad$  \\
$\varepsilon^{\mu\nu\lambda\rho}{P^{*}}_{\mu}u_{\nu}u^{\sigma}u_{\lambda}D_{\rho}{P^{*\dag}}_{\sigma}+\mathrm{H.c.}$ & 12 & $-2d^{(3)}_{2}-3d^{(3)}_{5}$ & $\frac{3}{2}\tilde{c}^{(3)}_{1}+\tilde{c}^{(3)}_{10}$ & 12 & $-D^{(3)}_{9}$  & $\frac{1}{2}\tilde{C}^{(3)}_{1}$  \\
$\varepsilon^{\mu\nu\lambda\rho}{P^{*}}_{\mu}u_{\nu}u^{\sigma}u_{\sigma}D_{\lambda}{P^{*\dag}}_{\rho}+\mathrm{H.c.}$ & 13 & $-2d^{(3)}_{1}-d^{(3)}_{2}$ & $\mathrm{I}\quad$ & 13 & $*\quad$  & $\mathrm{I}\quad$  \\
$\varepsilon^{\mu\nu\lambda\rho}{P^{*}}_{\mu}u^{\sigma}u_{\nu}u_{\lambda}D_{\rho}{P^{*\dag}}_{\sigma}+\mathrm{H.c.}$ &  &  &  & 14 & $*\quad$  & $-\tilde{C}^{(3)}_{1}-\tilde{C}^{(3)}_{10}$  \\
$\varepsilon^{\mu\nu\lambda\rho}{P^{*}}_{\mu}u_{\nu}u^{\sigma}u^{\delta}D_{\lambda\sigma\delta}{P^{*\dag}}_{\rho}+\mathrm{H.c.}$ & 14 & $2d^{(3)}_{3}+d^{(3)}_{4}$ & $\mathrm{I}\quad$ & 15 & $*\quad$  & $\mathrm{I}\quad$  \\
$\varepsilon^{\mu\nu\lambda\rho}{P^{*}}_{\mu}\la u_{\nu}u_{\lambda}u^{\sigma}\ra D_{\rho}{P^{*\dag}}_{\sigma}+\mathrm{H.c.}$ &  &  &  & 16 & $3D^{(3)}_{10}$  & $-\frac{3}{2}\tilde{C}^{(3)}_{2}$  \\
$\varepsilon^{\mu\nu\lambda\rho}{P^{*}}_{\mu}\la u_{\nu}u^{\sigma}\ra u_{\lambda}D_{\rho}{P^{*\dag}}_{\sigma}+\mathrm{H.c.}$ &  &  &  & 17 & $-2D^{(3)}_{5}+D^{(3)}_{6}$  & $\mathrm{I}\quad$  \\
$\varepsilon^{\mu\nu\lambda\rho}{P^{*}}_{\mu}\la u_{\nu}u^{\sigma}\ra u_{\lambda}D_{\sigma}{P^{*\dag}}_{\rho}$ &  &  &  & 18 & $*\quad$  & $\mathrm{I}\quad$  \\
$\varepsilon^{\mu\nu\lambda\rho}{P^{*}}_{\mu}\la u_{\nu}u^{\sigma}\ra u_{\sigma}D_{\lambda}{P^{*\dag}}_{\rho}$ &  &  &  & 19 & $-2D^{(3)}_{5}+3D^{(3)}_{6}$  & $\mathrm{I}\quad$  \\
$\varepsilon^{\mu\nu\lambda\rho}{P^{*}}_{\mu}\la u_{\nu}u^{\sigma}\ra u^{\delta}D_{\lambda\sigma\delta}{P^{*\dag}}_{\rho}$ &  &  &  & 20 & $2D^{(3)}_{7}-3D^{(3)}_{8}$  & $\mathrm{I}\quad$  \\
$P^{*\mu}u_{\mu}{f_{-}}^{\nu\lambda}D_{\nu}{P^{*\dag}}_{\lambda}+\mathrm{H.c.}$ & 15 & $-2d^{(3)}_{18}$ & $\mathrm{I}\quad$ & 21 & $-2D^{(3)}_{25}$  & $\mathrm{I}\quad$  \\
$P^{*\mu}u^{\nu}{f_{-\mu}}^{\lambda}D_{\nu}{P^{*\dag}}_{\lambda}+\mathrm{H.c.}$ & 16 & $4d^{(3)}_{19}$ & $\mathrm{I}\quad$ & 22 & $4D^{(3)}_{26}$  & $\mathrm{I}\quad$  \\
$P^{*\mu}u^{\nu}{f_{-\mu}}^{\lambda}D_{\lambda}{P^{*\dag}}_{\nu}+\mathrm{H.c.}$ & 17 & $-2d^{(3)}_{18}$ & $\tilde{c}^{(3)}_{15}$ & 23 & $-2D^{(3)}_{25}$  & $\tilde{C}^{(3)}_{21}$  \\
$P^{*\mu}u^{\nu}{f_{-\nu}}^{\lambda}D_{\lambda}{P^{*\dag}}_{\mu}+\mathrm{H.c.}$ & 18 & $2d^{(3)}_{17}$ & $-\tilde{c}^{(3)}_{2}$ & 24 & $2D^{(3)}_{24}$  & $-\tilde{C}^{(3)}_{3}$  \\
$P^{*\mu}u_{\mu}h^{\nu\lambda}D_{\nu}{P^{*\dag}}_{\lambda}+\mathrm{H.c.}$ & 19 & $2d^{(3)}_{22}$ & $\mathrm{I}\quad$ & 25 & $2D^{(3)}_{29}$  & $\mathrm{I}\quad$  \\
$P^{*\mu}u^{\nu}{h_{\mu}}^{\lambda}D_{\nu}{P^{*\dag}}_{\lambda}+\mathrm{H.c.}$ & 20 & $0\quad$ & $0\quad$ & 26 & $0\quad$  & $0\quad$  \\
$P^{*\mu}u^{\nu}{h_{\mu}}^{\lambda}D_{\lambda}{P^{*\dag}}_{\nu}+\mathrm{H.c.}$ & 21 & $-2d^{(3)}_{22}$ & $-\tilde{c}^{(3)}_{19}$ & 27 & $-2D^{(3)}_{29}$  & $-\tilde{C}^{(3)}_{25}$  \\
$P^{*\mu}u^{\nu}{h_{\nu}}^{\lambda}D_{\lambda}{P^{*\dag}}_{\mu}+\mathrm{H.c.}$ & 22 & $2d^{(3)}_{20}$ & $-\tilde{c}^{(3)}_{3}$ & 28 & $2D^{(3)}_{27}$  & $-\tilde{C}^{(3)}_{4}$  \\
$P^{*\mu}u^{\nu}h^{\lambda\rho}D_{\nu\lambda\rho}{P^{*\dag}}_{\mu}+\mathrm{H.c.}$ & 23 & $-2d^{(3)}_{21}$ & $-\tilde{c}^{(3)}_{4}$ & 29 & $-2D^{(3)}_{28}$  & $-\tilde{C}^{(3)}_{5}$  \\
$P^{*\mu}\la u_{\mu}{f_{-}}^{\nu\lambda}\ra D_{\nu}{P^{*\dag}}_{\lambda}+\mathrm{H.c.}$ &  &  &  & 30 & $-2D^{(3)}_{30}$  & $\mathrm{I}\quad$  \\
$P^{*\mu}\la u^{\nu}{f_{-\mu}}^{\lambda}\ra D_{\nu}{P^{*\dag}}_{\lambda}$ &  &  &  & 31 & $4D^{(3)}_{31}$  & $\mathrm{I}\quad$  \\
$P^{*\mu}\la u_{\mu}h^{\nu\lambda}\ra D_{\nu}{P^{*\dag}}_{\lambda}+\mathrm{H.c.}$ &  &  &  & 32 & $2D^{(3)}_{32}$  & $\mathrm{I}\quad$  \\
$\varepsilon^{\mu\nu\lambda\rho}{P^{*}}_{\mu}\nabla_{\nu}{f_{-\lambda}}^{\sigma}D_{\rho}{P^{*\dag}}_{\sigma}+\mathrm{H.c.}$ & 24 & $2d^{(3)}_{9}$ & $\mathrm{I}\quad$ & 33 & $2D^{(3)}_{15}$  & $\mathrm{I}\quad$  \\
$\varepsilon^{\mu\nu\lambda\rho}{P^{*}}_{\mu}\nabla_{\nu}{f_{-\lambda}}^{\sigma}D_{\sigma}{P^{*\dag}}_{\rho}$ & 25 & $-2d^{(3)}_{9}+2d^{(3)}_{10}$ & $\mathrm{I}\quad$ & 34 & $-2D^{(3)}_{15}+2D^{(3)}_{16}$  & $\mathrm{I}\quad$  \\
$\varepsilon^{\mu\nu\lambda\rho}{P^{*}}_{\mu}\nabla_{\nu}h^{\sigma\delta}D_{\lambda\sigma\delta}{P^{*\dag}}_{\rho}$ & 26 & $2d^{(3)}_{11}$ & $\mathrm{I}\quad$ & 35 & $2D^{(3)}_{17}$  & $\mathrm{I}\quad$  \\
$i\varepsilon^{\mu\nu\lambda\rho}{P^{*}}_{\mu}f_{+\nu\lambda}u^{\sigma}D_{\rho}{P^{*\dag}}_{\sigma}+\mathrm{H.c.}$ & 27 & $-2d^{(3)}_{12}$ & $-\tilde{c}^{(3)}_{5}$ & 36 & $-2D^{(3)}_{18}$  & $-\tilde{C}^{(3)}_{6}$  \\
$i\varepsilon^{\mu\nu\lambda\rho}{P^{*}}_{\mu}f_{+\nu\lambda}u^{\sigma}D_{\sigma}{P^{*\dag}}_{\rho}+\mathrm{H.c.}$ & 28 & $-d^{(3)}_{7}$ & $\mathrm{I}\quad$ & 37 & $-D^{(3)}_{12}$  & $\mathrm{I}\quad$  \\
$i\varepsilon^{\mu\nu\lambda\rho}{P^{*}}_{\mu}{f_{+\nu}}^{\sigma}u_{\lambda}D_{\rho}{P^{*\dag}}_{\sigma}+\mathrm{H.c.}$ & 29 & $4d^{(3)}_{12}$ & $2\tilde{c}^{(3)}_{5}$ & 38 & $4D^{(3)}_{18}$  & $2\tilde{C}^{(3)}_{6}$  \\
$i\varepsilon^{\mu\nu\lambda\rho}{P^{*}}_{\mu}{f_{+\nu}}^{\sigma}u_{\sigma}D_{\lambda}{P^{*\dag}}_{\rho}+\mathrm{H.c.}$ & 30 & $2d^{(3)}_{6}$ & $\mathrm{I}\quad$ & 39 & $2D^{(3)}_{11}$  & $\mathrm{I}\quad$  \\
$i\varepsilon^{\mu\nu\lambda\rho}{P^{*}}_{\mu}\la f_{+\nu\lambda}\ra u^{\sigma}D_{\rho}{P^{*\dag}}_{\sigma}+\mathrm{H.c.}$ & 31 & $-d^{(3)}_{13}$ & $-\frac{1}{2}\tilde{c}^{(3)}_{6}$ &  &  &  \\
$i\varepsilon^{\mu\nu\lambda\rho}{P^{*}}_{\mu}\la{f_{+\nu}}^{\sigma}\ra u_{\lambda}D_{\rho}{P^{*\dag}}_{\sigma}+\mathrm{H.c.}$ & 32 & $2d^{(3)}_{13}$ & $\tilde{c}^{(3)}_{6}$ &  &  &  \\
$i\varepsilon^{\mu\nu\lambda\rho}{P^{*}}_{\mu}\la f_{+\nu\lambda}u^{\sigma}\ra D_{\rho}{P^{*\dag}}_{\sigma}+\mathrm{H.c.}$ &  &  &  & 40 & $-D^{(3)}_{19}$  & $-\frac{1}{2}\tilde{C}^{(3)}_{7}$  \\
$i\varepsilon^{\mu\nu\lambda\rho}{P^{*}}_{\mu}\la{f_{+\nu}}^{\sigma}u_{\lambda}\ra D_{\rho}{P^{*\dag}}_{\sigma}+\mathrm{H.c.}$ &  &  &  & 41 & $2D^{(3)}_{19}$  & $\tilde{C}^{(3)}_{7}$  \\
$iP^{*\mu}\nabla_{\mu}{f_{+}}^{\nu\lambda}D_{\nu}{P^{*\dag}}_{\lambda}+\mathrm{H.c.}$ & 33 & $0\quad$ & $0\quad$ & 42 & $0\quad$  & $0\quad$  \\
$iP^{*\mu}\nabla^{\nu}{f_{+\nu}}^{\lambda}D_{\lambda}{P^{*\dag}}_{\mu}$ & 34 & $-2d^{(3)}_{23}$ & $-\tilde{c}^{(3)}_{7}$ & 43 & $-2D^{(3)}_{33}$  & $-\tilde{C}^{(3)}_{8}$  \\
$iP^{*\mu}\la\nabla_{\mu}{f_{+}}^{\nu\lambda}\ra D_{\nu}{P^{*\dag}}_{\lambda}+\mathrm{H.c.}$ & 35 & $0\quad$ & $0\quad$ &  &  &  \\
$iP^{*\mu}\la\nabla^{\nu}{f_{+\nu}}^{\lambda}\ra D_{\lambda}{P^{*\dag}}_{\mu}$ & 36 & $-2d^{(3)}_{24}$ & $-\tilde{c}^{(3)}_{8}$ &  &  &  \\
$\varepsilon^{\mu\nu\lambda\rho}{P^{*}}_{\mu}u_{\nu}\chip D_{\lambda}{P^{*\dag}}_{\rho}+\mathrm{H.c.}$ & 37 & $-2d^{(3)}_{14}$ & $\mathrm{I}\quad$ & 44 & $-2D^{(3)}_{20}$  & $\mathrm{I}\quad$  \\
$\varepsilon^{\mu\nu\lambda\rho}{P^{*}}_{\mu}\la u_{\nu}\chip\ra D_{\lambda}{P^{*\dag}}_{\rho}$ & 38 & $-2d^{(3)}_{15}$ & $\mathrm{I}\quad$ & 45 & $-2D^{(3)}_{21}$  & $\mathrm{I}\quad$  \\
$\varepsilon^{\mu\nu\lambda\rho}{P^{*}}_{\mu}\la\chip\ra u_{\nu}D_{\lambda}{P^{*\dag}}_{\rho}$ &  &  &  & 46 & $-2D^{(3)}_{22}$  & $\mathrm{I}\quad$  \\
$iP^{*\mu}u^{\nu}\chim D_{\nu}{P^{*\dag}}_{\mu}+\mathrm{H.c.}$ & 39 & $-2d^{(3)}_{16}$ & $-\tilde{c}^{(3)}_{9}$ & 47 & $-2D^{(3)}_{23}$  & $-\tilde{C}^{(3)}_{9}$  \\
$i\varepsilon^{\mu\nu\lambda\rho}{P^{*}}_{\mu}\nabla_{\nu}\chi_{-} D_{\lambda}{P^{*\dag}}_{\rho}$ & 40 & $-2d^{(3)}_{25}$ & $\mathrm{I}\quad$ & 48 & $-2D^{(3)}_{13}$  & $\mathrm{I}\quad$  \\
$i\varepsilon^{\mu\nu\lambda\rho}{P^{*}}_{\mu}\la\nabla_{\nu}\chi_{-}\ra D_{\lambda}{P^{*\dag}}_{\rho}$ & 41 & $-2d^{(3)}_{8}$ & $\mathrm{I}\quad$ & 49 & $-2D^{(3)}_{14}$  & $\mathrm{I}\quad$  \\
$P u^{\mu}u_{\mu}u^{\nu}{P^{*\dag}}_{\nu}+\mathrm{H.c.}$ & 42 & $4d^{(3)}_{1}$ & $\tilde{c}^{(3)}_{10}-2\tilde{c}^{(3)}_{13}$ & 50 & $2D^{(3)}_{1}$  & $*\quad$  \\
$P u^{\mu}u^{\nu}u_{\mu}{P^{*\dag}}_{\nu}+\mathrm{H.c.}$ & 43 & $2d^{(3)}_{2}$ & $-\tilde{c}^{(3)}_{10}$ & 51 & $2D^{(3)}_{2}$  & $*\quad$  \\
$P u^{\mu}u^{\nu}u_{\nu}{P^{*\dag}}_{\mu}+\mathrm{H.c.}$ &  &  &  & 52 & $2D^{(3)}_{1}$  & $*\quad$  \\
$P u^{\mu}u^{\nu}u^{\lambda}D_{\mu\nu}{P^{*\dag}}_{\lambda}+\mathrm{H.c.}$ & 44 & $-4d^{(3)}_{3}-2d^{(3)}_{11}$ & $*\quad$ & 53 & $*\quad$  & $*\quad$  \\
$P u^{\mu}u^{\nu}u^{\lambda}D_{\mu\lambda}{P^{*\dag}}_{\nu}+\mathrm{H.c.}$ & 45 & $-2d^{(3)}_{4}+2d^{(3)}_{11}$ & $*\quad$ & 54 & $*\quad$  & $*\quad$  \\
$P u^{\mu}u^{\nu}u^{\lambda}D_{\nu\lambda}{P^{*\dag}}_{\mu}+\mathrm{H.c.}$ &  &  &  & 55 & $*\quad$  & $*\quad$  \\
$P\la u^{\mu}u_{\mu}\ra u^{\nu}{P^{*\dag}}_{\nu}+\mathrm{H.c.}$ &  &  &  & 56 & $2D^{(3)}_{5}$  & $-\frac{3}{2}\tilde{C}^{(3)}_{17}+\frac{1}{2}\tilde{C}^{(3)}_{19}$  \\
$P\la u^{\mu}u_{\mu}u^{\nu}\ra{P^{*\dag}}_{\nu}+\mathrm{H.c.}$ &  &  &  & 57 & $2D^{(3)}_{6}$  & $-\tilde{C}^{(3)}_{17}+\tilde{C}^{(3)}_{19}$  \\
$P\la u^{\mu}u^{\nu}\ra u^{\lambda}D_{\mu\nu}{P^{*\dag}}_{\lambda}+\mathrm{H.c.}$ &  &  &  & 58 & $D^{(3)}_{7}$  & $*\quad$  \\
$P\la u^{\mu}u^{\nu}u^{\lambda}\ra D_{\mu\nu}{P^{*\dag}}_{\lambda}+\mathrm{H.c.}$ &  &  &  & 59 & $2D^{(3)}_{7}-2D^{(3)}_{8}$  & $*\quad$  \\
$\varepsilon^{\mu\nu\lambda\rho}P u_{\mu}f_{-\nu\lambda}{P^{*\dag}}_{\rho}+\mathrm{H.c.}$ & 46 & $-2d^{(3)}_{19}$ & $-\frac{1}{2}\tilde{c}^{(3)}_{16}$ & 60 & $-2D^{(3)}_{26}$  & $-\frac{1}{2}\tilde{C}^{(3)}_{22}$  \\
$\varepsilon^{\mu\nu\lambda\rho}P f_{-\mu\nu}u_{\lambda}{P^{*\dag}}_{\rho}+\mathrm{H.c.}$ & 47 & $-d^{(3)}_{18}$ & $\frac{1}{2}\tilde{c}^{(3)}_{15}$ & 61 & $-D^{(3)}_{25}$  & $\frac{1}{2}\tilde{C}^{(3)}_{21}$  \\
$\varepsilon^{\mu\nu\lambda\rho}P u_{\mu}{f_{-\nu}}^{\sigma}D_{\lambda\sigma}{P^{*\dag}}_{\rho}+\mathrm{H.c.}$ & 48 & $2d^{(3)}_{18}-4d^{(3)}_{19}$ & $-\tilde{c}^{(3)}_{15}-\tilde{c}^{(3)}_{16}$ & 62 & $2D^{(3)}_{25}-4D^{(3)}_{26}$  & $-\tilde{C}^{(3)}_{21}-\tilde{C}^{(3)}_{22}$  \\
$\varepsilon^{\mu\nu\lambda\rho}P f_{-\mu\nu}u^{\sigma}D_{\lambda\sigma}{P^{*\dag}}_{\rho}+\mathrm{H.c.}$ & 49 & $-d^{(3)}_{18}+2d^{(3)}_{19}$ & $\frac{1}{2}\tilde{c}^{(3)}_{15}+\frac{1}{2}\tilde{c}^{(3)}_{16}$ & 63 & $-D^{(3)}_{25}+2D^{(3)}_{26}$  & $\frac{1}{2}\tilde{C}^{(3)}_{21}+\frac{1}{2}\tilde{C}^{(3)}_{22}$  \\
$\varepsilon^{\mu\nu\lambda\rho}P u_{\mu}{h_{\nu}}^{\sigma}D_{\lambda\sigma}{P^{*\dag}}_{\rho}+\mathrm{H.c.}$ & 50 & $2d^{(3)}_{22}$ & $\tilde{c}^{(3)}_{19}$ & 64 & $2D^{(3)}_{29}$  & $\tilde{C}^{(3)}_{25}$  \\
$\varepsilon^{\mu\nu\lambda\rho}P{h_{\mu}}^{\sigma}u_{\nu}D_{\lambda\sigma}{P^{*\dag}}_{\rho}+\mathrm{H.c.}$ & 51 & $-2d^{(3)}_{22}$ & $-\tilde{c}^{(3)}_{19}$ & 65 & $-2D^{(3)}_{29}$  & $-\tilde{C}^{(3)}_{25}$  \\
$\varepsilon^{\mu\nu\lambda\rho}P\la u_{\mu}f_{-\nu\lambda}\ra{P^{*\dag}}_{\rho}+\mathrm{H.c.}$ &  &  &  & 66 & $-2D^{(3)}_{31}$  & $-\frac{1}{2}\tilde{C}^{(3)}_{31}$  \\
$\varepsilon^{\mu\nu\lambda\rho}P\la u_{\mu}{f_{-\nu}}^{\sigma}\ra D_{\lambda\sigma}{P^{*\dag}}_{\rho}+\mathrm{H.c.}$ &  &  &  & 67 & $2D^{(3)}_{30}-4D^{(3)}_{31}$  & $-\tilde{C}^{(3)}_{30}-\tilde{C}^{(3)}_{31}$  \\
$\varepsilon^{\mu\nu\lambda\rho}P\la u_{\mu}{h_{\nu}}^{\sigma}\ra D_{\lambda\sigma}{P^{*\dag}}_{\rho}+\mathrm{H.c.}$ &  &  &  & 68 & $2D^{(3)}_{32}$  & $\tilde{C}^{(3)}_{32}$  \\
$P\nabla^{\mu}{f_{-\mu}}^{\nu}{P^{*\dag}}_{\nu}+\mathrm{H.c.}$ & 52 & $2d^{(3)}_{9}$ & $\tilde{c}^{(3)}_{24}$ & 69 & $2D^{(3)}_{15}$  & $\tilde{C}^{(3)}_{33}$  \\
$P\nabla^{\mu}{f_{-}}^{\nu\lambda}D_{\mu\nu}{P^{*\dag}}_{\lambda}+\mathrm{H.c.}$ & 53 & $2d^{(3)}_{10}-2d^{(3)}_{11}$ & $*\quad$ & 70 & $2D^{(3)}_{16}-2D^{(3)}_{17}$  & $*\quad$  \\
$P\nabla^{\mu}h^{\nu\lambda}D_{\mu\nu}{P^{*\dag}}_{\lambda}+\mathrm{H.c.}$ & 54 & $-2d^{(3)}_{11}$ & $-\tilde{c}^{(3)}_{26}$ & 71 & $-2D^{(3)}_{17}$  & $-\tilde{C}^{(3)}_{35}$  \\
$iP{f_{+}}^{\mu\nu}u_{\mu}{P^{*\dag}}_{\nu}+\mathrm{H.c.}$ & 55 & $2d^{(3)}_{6}$ & $\tilde{c}^{(3)}_{30}$ & 72 & $2D^{(3)}_{11}$  & $\tilde{C}^{(3)}_{39}$  \\
$iP u^{\mu}{f_{+\mu}}^{\nu}{P^{*\dag}}_{\nu}+\mathrm{H.c.}$ & 56 & $-2d^{(3)}_{6}$ & $-\tilde{c}^{(3)}_{30}$ & 73 & $-2D^{(3)}_{11}$  & $-\tilde{C}^{(3)}_{39}$  \\
$iP{f_{+}}^{\mu\nu}u^{\lambda}D_{\mu\lambda}{P^{*\dag}}_{\nu}+\mathrm{H.c.}$ & 57 & $2d^{(3)}_{7}-2d^{(3)}_{11}$ & $-\tilde{c}^{(3)}_{26}-2\tilde{c}^{(3)}_{28}$ & 74 & $2D^{(3)}_{12}-2D^{(3)}_{17}$  & $-\tilde{C}^{(3)}_{35}-2\tilde{C}^{(3)}_{37}$  \\
$iP u^{\mu}{f_{+}}^{\nu\lambda}D_{\mu\nu}{P^{*\dag}}_{\lambda}+\mathrm{H.c.}$ & 58 & $-2d^{(3)}_{7}+2d^{(3)}_{11}$ & $\tilde{c}^{(3)}_{26}+2\tilde{c}^{(3)}_{28}$ & 75 & $-2D^{(3)}_{12}+2D^{(3)}_{17}$  & $\tilde{C}^{(3)}_{35}+2\tilde{C}^{(3)}_{37}$  \\
$iP\la{f_{+}}^{\mu\nu}\ra u_{\mu}{P^{*\dag}}_{\nu}+\mathrm{H.c.}$ & 59 & $0\quad$ & $0\quad$ &  &  &  \\
$iP\la{f_{+}}^{\mu\nu}\ra u^{\lambda}D_{\mu\lambda}{P^{*\dag}}_{\nu}+\mathrm{H.c.}$ & 60 & $0\quad$ & $0\quad$ &  &  &  \\
$iP\la{f_{+}}^{\mu\nu}u_{\mu}\ra{P^{*\dag}}_{\nu}+\mathrm{H.c.}$ &  &  &  & 76 & $0\quad$  & $0\quad$  \\
$iP\la{f_{+}}^{\mu\nu}u^{\lambda}\ra D_{\mu\lambda}{P^{*\dag}}_{\nu}+\mathrm{H.c.}$ &  &  &  & 77 & $0\quad$  & $0\quad$  \\
$i\varepsilon^{\mu\nu\lambda\rho}P\nabla_{\mu}{f_{+\nu}}^{\sigma}D_{\lambda\sigma}{P^{*\dag}}_{\rho}+\mathrm{H.c.}$ & 61 & $0\quad$ & $0\quad$ & 78 & $0\quad$  & $0\quad$  \\
$i\varepsilon^{\mu\nu\lambda\rho}P\la\nabla_{\mu}{f_{+\nu}}^{\sigma}\ra D_{\lambda\sigma}{P^{*\dag}}_{\rho}+\mathrm{H.c.}$ & 62 & $0\quad$ & $0\quad$ &  &  &  \\
$P u^{\mu}\chip{P^{*\dag}}_{\mu}+\mathrm{H.c.}$ & 63 & $2d^{(3)}_{14}$ & $-\tilde{c}^{(3)}_{37}$ & 79 & $2D^{(3)}_{20}$  & $-\tilde{C}^{(3)}_{44}$  \\
$P\chip u^{\mu}{P^{*\dag}}_{\mu}+\mathrm{H.c.}$ & 64 & $2d^{(3)}_{14}$ & $-\tilde{c}^{(3)}_{37}$ & 80 & $2D^{(3)}_{20}$  & $-\tilde{C}^{(3)}_{44}$  \\
$P\la u^{\mu}\chip\ra{P^{*\dag}}_{\mu}+\mathrm{H.c.}$ & 65 & $2d^{(3)}_{15}$ & $-\tilde{c}^{(3)}_{38}$ & 81 & $2D^{(3)}_{21}$  & $-\tilde{C}^{(3)}_{45}$  \\
$P\la\chip\ra u^{\mu}{P^{*\dag}}_{\mu}+\mathrm{H.c.}$ &  &  &  & 82 & $2D^{(3)}_{22}$  & $-\tilde{C}^{(3)}_{46}$  \\
$iP\nabla^{\mu}\chi_{-}{P^{*\dag}}_{\mu}+\mathrm{H.c.}$ & 66 & $2d^{(3)}_{25}$ & $-\tilde{c}^{(3)}_{40}$ & 83 & $2D^{(3)}_{13}$  & $-\tilde{C}^{(3)}_{48}$  \\
$iP\la\nabla^{\mu}\chi_{-}\ra{P^{*\dag}}_{\mu}+\mathrm{H.c.}$ & 67 & $2d^{(3)}_{8}$ & $-\tilde{c}^{(3)}_{41}$ & 84 & $2D^{(3)}_{14}$  & $-\tilde{C}^{(3)}_{49}$  \\
\hline\hline
\end{longtable}

The long relations in the fourth column of Table \ref{rp3} are
\begin{eqnarray}
\tilde{c}^{(3)}_{44}&=&(\tilde{c}^{(3)}_{10}+2\tilde{c}^{(3)}_{11}-\tilde{c}^{(3)}_{26})-2\tilde{c}^{(3)}_{14},\nonumber\\
\tilde{c}^{(3)}_{45}&=&-(\tilde{c}^{(3)}_{10}+2\tilde{c}^{(3)}_{11}-\tilde{c}^{(3)}_{26}),\nonumber\\
\tilde{c}^{(3)}_{53}&=&\tilde{c}^{(3)}_{24}+\tilde{c}^{(3)}_{25}-\tilde{c}^{(3)}_{26}.
\end{eqnarray}

The long relations in the sixth column of Table \ref{rp3} are
\begin{align}
\begin{autobreak}
~
\tilde{C}^{(3)}_{10}=-D^{(3)}_{2}-D^{(3)}_{6}+D^{(3)}_{9},\;
\tilde{C}^{(3)}_{11}=D^{(3)}_{2}+D^{(3)}_{4}+D^{(3)}_{6}+D^{(3)}_{8},\;
\tilde{C}^{(3)}_{13}=-2D^{(3)}_{1}-D^{(3)}_{2}-3D^{(3)}_{6},\;
\tilde{C}^{(3)}_{14}=D^{(3)}_{2}+D^{(3)}_{6}+D^{(3)}_{9},\;
\tilde{C}^{(3)}_{15}=2D^{(3)}_{3}+D^{(3)}_{4}+3D^{(3)}_{8},\;
\tilde{C}^{(3)}_{18}=2D^{(3)}_{5}-D^{(3)}_{6}-D^{(3)}_{8},\;
\tilde{C}^{(3)}_{53}=\tilde{C}^{(3)}_{55}=-2D^{(3)}_{3}-2D^{(3)}_{7}-D^{(3)}_{17},\;
\tilde{C}^{(3)}_{54}=-2D^{(3)}_{4}-2D^{(3)}_{7}+2D^{(3)}_{17}.
\end{autobreak}
\end{align}
The long relations in the seventh column in Table \ref{rp3} are
\begin{align}
\begin{autobreak}
~
\tilde{C}^{(3)}_{50}=\tilde{C}^{(3)}_{52}=\frac{1}{2}\tilde{C}^{(3)}_{1}+\tilde{C}^{(3)}_{10}-\tilde{C}^{(3)}_{13}+\tilde{C}^{(3)}_{17}-\tilde{C}^{(3)}_{19},\;
\tilde{C}^{(3)}_{51}=-\tilde{C}^{(3)}_{1}-2\tilde{C}^{(3)}_{10}+\tilde{C}^{(3)}_{17}-\tilde{C}^{(3)}_{19},\;
\tilde{C}^{(3)}_{53}=\tilde{C}^{(3)}_{55}=\frac{1}{2}\tilde{C}^{(3)}_{1}+\tilde{C}^{(3)}_{10}+\tilde{C}^{(3)}_{11}-\tilde{C}^{(3)}_{15}+\tilde{C}^{(3)}_{17}+\tilde{C}^{(3)}_{18}-\tilde{C}^{(3)}_{20}-\frac{1}{2}\tilde{C}^{(3)}_{35},\;
\tilde{C}^{(3)}_{54}=-\tilde{C}^{(3)}_{1}-2\tilde{C}^{(3)}_{10}-2\tilde{C}^{(3)}_{11}+\tilde{C}^{(3)}_{17}+\tilde{C}^{(3)}_{18}-\tilde{C}^{(3)}_{20}+\tilde{C}^{(3)}_{35},\;
\tilde{C}^{(3)}_{58}=-\frac{3}{2}\tilde{C}^{(3)}_{17}-\frac{3}{2}\tilde{C}^{(3)}_{18}+\frac{1}{2}\tilde{C}^{(3)}_{20},\;
\tilde{C}^{(3)}_{59}=-\tilde{C}^{(3)}_{17}-\tilde{C}^{(3)}_{18}+\tilde{C}^{(3)}_{20},\;
\tilde{C}^{(3)}_{70}=\tilde{C}^{(3)}_{33}+\tilde{C}^{(3)}_{34}-\tilde{C}^{(3)}_{35}.
\end{autobreak}
\end{align}

\begin{table*}[!h]
\caption{\label{hp3}The $\mathcal{O}(p^3)$ order results in the HQ limit. When a term $P_n$ is not given a label in the 2nd or 5th (3rd or 6th) column, it is not independent and can be expressed with terms having a label in the 2nd and 5th (3rd and 6th) columns.}
%\begin{ruledtabular}
\begin{tabular}{lcclcc}
\hline\hline $P_n$ & $SU(2)$ & $SU(3)$ & $P_n$ & $SU(2)$ & $SU(3)$\\
\hline%\addlinespace[2pt]
$\la H u^{\mu}u_{\mu}u^{\nu}\gamf\gamma_{\nu}\Hb\ra+\mathrm{H.c.}$ & 1 & 1 & $\varepsilon^{\mu\nu\lambda\rho}\la H\la f_{+\mu\nu}\ra u_{\lambda}v_{\rho}\Hb\ra$ & 13 &  \\
$\la H u^{\mu}u^{\nu}u_{\mu}\gamf\gamma_{\nu}\Hb\ra$ & 2 & 2 & $\varepsilon^{\mu\nu\lambda\rho}\la H\la f_{+\mu\nu}u_{\lambda}\ra v_{\rho}\Hb\ra$ &  & 19  \\
$\la H u^{\mu}u^{\nu}u^{\lambda}\gamf\gamma_{\mu}v_{\nu}v_{\lambda}\Hb\ra+\mathrm{H.c.}$ & 3 & 3 & $\la H u^{\mu}\chip\gamf\gamma_{\mu}\Hb\ra+\mathrm{H.c.}$ & 14 & 20  \\
$\la H u^{\mu}u^{\nu}u^{\lambda}\gamf\gamma_{\nu}v_{\mu}v_{\lambda}\Hb\ra$ & 4 & 4 & $\la H\la u^{\mu}\chip\ra\gamf\gamma_{\mu}\Hb\ra$ & 15 & 21  \\
$\la H\la u^{\mu}u_{\mu}\ra u^{\nu}\gamf\gamma_{\nu}\Hb\ra$ &  & 5 & $\la H\la\chip\ra u^{\mu}\gamf\gamma_{\mu}\Hb\ra$ &  & 22  \\
$\la H\la u^{\mu}u_{\mu}u^{\nu}\ra\gamf\gamma_{\nu}\Hb\ra$ &  & 6 & $\la H u^{\mu}\chim v_{\mu}\Hb\ra+\mathrm{H.c.}$ & 16 & 23  \\
$\la H\la u^{\mu}u^{\nu}\ra u^{\lambda}\gamf\gamma_{\mu}v_{\nu}v_{\lambda}\Hb\ra$ &  & 7 & $i\la H u^{\mu}{f_{-\mu}}^{\nu}v_{\nu}\Hb\ra+\mathrm{H.c.}$ & 17 & 24  \\
\hline\hline
\end{tabular}
%\end{ruledtabular}
\end{table*}

\begin{longtable}{lcrcrlcrcr}
\caption{The $\mathcal{O}(p^4)$ order relativistic results. The columns 2, 3, 7, and 8 (4, 5, 9, and 10) are for the flavor $SU(2)$ (SU(3)) case. When a term $O_n$ is not given a label in the 2nd or 7th (4th or 8th) column, it is not independent and can be expressed with terms having a label in the $SU(2)$ ($SU(3)$) case. ``I'' means that the structures of those terms are chosen as independent ones in the HQ limit. ``P.'' stands for parity-transformed part.}\label{rp4}\\
\hline\hline $O_n$ & $SU(2)$ & $\tilde{c}^{(4)}_n$ & $SU(3)$ & $\tilde{C}^{(4)}_n$ & $O_n$ & $SU(2)$ & $\tilde{c}^{(4)}_n$ & $SU(3)$ & $\tilde{C}^{(4)}_n$\\
\hline\endfirsthead

\hline\hline $O_n$ & $SU(2)$ & $\tilde{c}^{(4)}_n$ & $SU(3)$ & $\tilde{C}^{(4)}_n$ & $O_n$ & $SU(2)$ & $\tilde{c}^{(4)}_n$ & $SU(3)$ & $\tilde{C}^{(4)}_n$\\
\hline\endhead

\hline % \multicolumn{8}{c}{Continued on next page}\\
\endfoot

\hline\endlastfoot

$P\la u^{\mu}u_{\mu}\ra u^{\nu}u_{\nu}P^{\dag}$ & 1& $\mathrm{I}\quad$ & 1& $\mathrm{I}\quad$ & $iP^{*\mu}\nabla_{\mu}\nabla^{\nu}{f_{+}}^{\lambda\rho}D_{\nu\lambda}{P^{*\dag}}_{\rho}+\mathrm{H.c.}$ & 182& $\mathrm{I}\quad$ & 334& $\mathrm{I}\quad$  \\
$P\la u^{\mu}u^{\nu}\ra u_{\mu}u_{\nu}P^{\dag}$ & 2& $\mathrm{I}\quad$ & 2& $\mathrm{I}\quad$ & $iP^{*\mu}\nabla^{\nu}\nabla^{\lambda}{f_{+\mu}}^{\rho}D_{\nu\lambda}{P^{*\dag}}_{\rho}$ & 183& $\mathrm{I}\quad$ & 335& $\mathrm{I}\quad$  \\
$P\la u^{\mu}u_{\mu}\ra u^{\nu}u^{\lambda}D_{\nu\lambda}P^{\dag}$ & 3& $\mathrm{I}\quad$ & 3& $\mathrm{I}\quad$ & $P^{*\mu}{f_{+\mu}}^{\nu}{f_{+\nu}}^{\lambda}{P^{*\dag}}_{\lambda}$ & 184& $\mathrm{I}\quad$ & 336& $\mathrm{I}\quad$  \\
$P\la u^{\mu}u^{\nu}\ra u_{\mu}u^{\lambda}D_{\nu\lambda}P^{\dag}+\mathrm{H.c.}$ & 4& $\mathrm{I}\quad$ & 4& $\mathrm{I}\quad$ & $P^{*\mu}{f_{+}}^{\nu\lambda}f_{+\mu\nu}{P^{*\dag}}_{\lambda}$ & 185&  & 337&  \\
$P\la u^{\mu}u^{\nu}\ra u^{\lambda}u_{\lambda}D_{\mu\nu}P^{\dag}$ &  &  & 5& $\mathrm{I}\quad$ & $P^{*\mu}{f_{+}}^{\nu\lambda}f_{+\nu\lambda}{P^{*\dag}}_{\mu}$ & 186&  & 338&  \\
$P\la u^{\mu}u^{\nu}\ra u^{\lambda}u^{\rho}D_{\mu\nu\lambda\rho}P^{\dag}$ & 5& $\mathrm{I}\quad$ & 6& $\mathrm{I}\quad$ & $P^{*\mu}{f_{+\mu}}^{\nu}{f_{+}}^{\lambda\rho}D_{\nu\lambda}{P^{*\dag}}_{\rho}$ & 187& $\mathrm{I}\quad$ & 339& $\mathrm{I}\quad$  \\
$P\la u^{\mu}u_{\mu}u^{\nu}\ra u_{\nu}P^{\dag}$ &  &  & 7& $\mathrm{I}\quad$ & $P^{*\mu}{f_{+}}^{\nu\lambda}{f_{+\mu}}^{\rho}D_{\nu\rho}{P^{*\dag}}_{\lambda}$ & 188&  & 340&  \\
$P\la u^{\mu}u_{\mu}u^{\nu}\ra u^{\lambda}D_{\nu\lambda}P^{\dag}$ &  &  & 8& $\mathrm{I}\quad$ & $P^{*\mu}{f_{+}}^{\nu\lambda}{f_{+\nu}}^{\rho}D_{\lambda\rho}{P^{*\dag}}_{\mu}$ & 189&  & 341&  \\
\hline\hline
\end{longtable}

\begin{longtable}{lcclcc}
\caption{The $\mathcal{O}(p^4)$ order results in the HQ limit. When a term $P_n$ is not given a label in the 2nd or 5th (3rd or 6th) column, it is not independent and can be expressed with terms having a label in the 2nd and 5th (3rd and 6th) columns.}\label{hp4}\\
\hline\hline $P_n$ & $SU(2)$ & $SU(3)$ & $P_n$ & $SU(2)$ & $SU(3)$\\
\hline\endfirsthead

\hline\hline $P_n$ & $SU(2)$ & $SU(3)$ & $P_n$ & $SU(2)$ & $SU(3)$\\
\hline\endhead

\hline\hline % \multicolumn{8}{c}{Continued on next page}\\
\endfoot

\hline\endlastfoot

$\la H\la u^{\mu}u_{\mu}\ra u^{\nu}u_{\nu}\Hb\ra$ & 1 & 1 & $\la H\la{f_{+}}^{\mu\nu}u^{\lambda}\ra u^{\rho}\sigma_{\mu\rho}v_{\nu}v_{\lambda}\Hb\ra$ & 72 &  \\
$\la H\la u^{\mu}u^{\nu}\ra u_{\mu}u_{\nu}\Hb\ra$ & 2 & 2 & $\la H\nabla^{\mu}\nabla_{\mu}\chi_{+}\Hb\ra$ &  & 111  \\
$\la H\la u^{\mu}u_{\mu}\ra u^{\nu}u^{\lambda}v_{\nu}v_{\lambda}\Hb\ra$ & 3 & 3 & $\la H\nabla^{\mu}\nabla^{\nu}\chi_{+} v_{\mu}v_{\nu}\Hb\ra$ &  & 112  \\
$\la H\la u^{\mu}u^{\nu}\ra u_{\mu}u^{\lambda}v_{\nu}v_{\lambda}\Hb\ra+\mathrm{H.c.}$ & 4 & 4 & $\la H\la\nabla^{\mu}\nabla_{\mu}\chi_{+}\ra\Hb\ra$ &  & 113  \\
$\la H\la u^{\mu}u^{\nu}\ra u^{\lambda}u_{\lambda}v_{\mu}v_{\nu}\Hb\ra$ &  & 5 & $\la H\la\nabla^{\mu}\nabla^{\nu}\chi_{+}\ra v_{\mu}v_{\nu}\Hb\ra$ &  & 114  \\
$\la H\la u^{\mu}u^{\nu}\ra u^{\lambda}u^{\rho}v_{\mu}v_{\nu}v_{\lambda}v_{\rho}\Hb\ra$ & 5 & 6 & $\la H\la{f_{+}}^{\mu\nu}u_{\mu}u^{\lambda}\ra\sigma_{\nu\lambda}\Hb\ra+\mathrm{H.c.}$ &  & 115  \\
$\la H\la u^{\mu}u_{\mu}u^{\nu}\ra u_{\nu}\Hb\ra$ &  & 7 & $\la H\la{f_{+}}^{\mu\nu}u^{\lambda}u_{\lambda}\ra\sigma_{\mu\nu}\Hb\ra$ &  & 116  \\
$\la H\la u^{\mu}u_{\mu}u^{\nu}\ra u^{\lambda}v_{\nu}v_{\lambda}\Hb\ra$ &  & 8 & $\la H\la{f_{+}}^{\mu\nu}u^{\lambda}u^{\rho}\ra\sigma_{\mu\nu}v_{\lambda}v_{\rho}\Hb\ra$ &  & 117  \\
\hline\hline
\end{longtable}

\bibliography{references}% Produces the bibliography via BibTeX.
\end{document}